\documentclass{JHEP3}
\usepackage{graphics}
\usepackage{amssymb,amsfonts,epsfig,euscript,array,cite,amsmath}

\newcounter{multieqs}

\newcommand{\be}{\begin{equation}}
\newcommand{\ee}{\end{equation}}
\newcommand{\eq}[1]{(\ref{#1})}

\newcommand{\bm}[1]{\mbox{\boldmath $#1$}}

\def\nn{\nonumber}
\def\bea{\begin{eqnarray}}
\def\eea{\end{eqnarray}}
\let\bm=\bibitem

\def\beqa{\begin{eqnarray}}
\def\eeqa{\end{eqnarray}}
\def\beq{\begin{equation}}
\def\eeq{\end{equation}}

\def\one{\mbox{1 \kern-.59em {\rm l}}}

\def\a{\alpha}      
\def\b{\beta}       
    
\def\d{\delta}  \def\D{\Delta}

\def\m{\mu} 
\def\o{\omega}

\def\s{\sigma}  
\def\t{\tau}

\def\cD{{\cal D}}

\def\ss {{\s \hspace{-6.4pt} \slash}\;}

\def\Jh{\hat{J}}

\def\omh{\hat{\omega}}

\def\Phit{\tilde{\Phi}}

\def\Vt{\tilde{V}}
\def\Xt{\tilde{X}}

\def\ft{\tilde{f}}

\def\sr{{\sf r}}
\def\ss{{\sf s}}

\def\vP{\vec{P}}
\def\vS{\vec{S}}
\def\vX{\vec{X}}
\def\vY{\vec{Y}}
\def\vn{\vec{n}}
\def\vp{\vec{p}}
\def\vx{\vec{x}}
\def\vy{\vec{y}}

\def\d{\delta}\def\D{\Delta}

 \def\del{\partial}

\def\uno{\mbox{1 \kern-.59em {\rm l}}}

\def\dag{{^{\dagger}}}

\def\one{1\!\!1\,\,}

\def\vac{|0\rangle}
\def\lvac{\langle 0|}

\def\bcomment#1{}

\newcommand{\third}{$3^\text{rd} \ $}

\title{Local Commutativity and Causality in Interacting PP-wave 
String Field Theory}

\author{Chong-Sun Chu \\
Department of Physics, National Tsing Hua University,
Hsinchu, Taiwan 300, R.O.C. \\
Centre for Particle Theory, Department of Mathematical Sciences,
University of Durham, Durham, DH1 3LE, United Kingdom \\
Email: \email{Chong-Sun.Chu@durham.ac.uk} }
\author{Konstantinos Kyritsis\\
Centre for Particle Theory, Department of Mathematical Sciences,
University of Durham, Durham, DH1 3LE, United Kingdom \\
Email: \email{Konstantinos.Kyritsis@durham.ac.uk } }

\abstract{
In this paper, we 
extend our previous study of causality and local commutativity of
string fields in the pp-wave lightcone string field theory to include
interaction. Contrary to the flat space case result of Lowe, Polchinski,
Susskind, Thorlacius and Uglum, we found that the pp-wave interaction
does not affect the local commutativity condition. Our results show that
the pp-wave lightcone string field theory is not continuously
connected with the flat space one.
We also discuss the relation between  the condition of 
local commutativity and causality.  While the two notions are closely
related in a point particle theory, 
their relation is less clear in string theory.
We suggest that string local commutativity may be relevant
for an operational defintion of causality using strings as probes.
} 

\keywords{local commutativity, causality, string field theory, pp-wave } 
\preprint{{\tt hep-th/0407046}}

\begin{document}


\section{Introduction}

Perturbative string theory provides a consistent quantization of
gravity in the background of a flat spacetime. It is generally 
believed that nonperturbative string theory will lead to a
fully consistent theory of quantum gravity and allows us to answer
important questions concerning gravity such as the statistical 
origin of the blackhole entropy, 
the holographic nature of gravity, the blackhole information paradox 
and the fate of geometry at the very short distance scale etc. Over the years,
many different frameworks have been proposed and studied. The most notably
ones are the string field theory approach (lightcone string field theory
\cite{sft1,sft2,sft3,sft4},
Witten covariant string field theory \cite{witten}), the Matrix model approach 
(BFSS matrix model \cite{BFSS}, IKKT matrix model \cite{IKKT}) 
and the gauge/gravity approach of Maldacena\cite{mal}. 
Although it looks  quite promising, we have not 
understood well enough of these different frameworks to 
obtain a  controllable  background
independent nonperturbative formulation of string theory. It is
therefore of great interest to understand better and to further
develop these (and others, possibly new ones) formulations, 
and to understand the possible relations among them. 

A common characteristic of these formulations
is that they are formulated on objects that may not be directly related
to the observables of interests (e.g. in AdS/CFT proposal, the bulk
physics are dual to the quantities defined on the boundary; in 
string field theory, the string field is not directly related to the 
S-matrix). Thus it is possible that certain basic physical 
requirements which are expected to hold 
for any physical theory may not be so apparent and needed to be
examined in more details. One of these is the requirement of
causality.   

Causality  is easy to formulate in a point particle theory in terms of
the propagation of light signal.  It is also easy to implement in 
the classical theory, which amounts to the imposition of an
appropriate boundary condition on the Green function. 
In a quantum field theory of point particle, causality is guaranteed
by, among others, the condition of {\it local commutativity} 
(also called microscopic
causality) of quantum fields \cite{PCT}: quantum fields at spacelike 
separation (anti)commute. Now string is extended and is nonlocal, 
it is a priori not clear whether the theory is causal. 
Causality in the AdS/CFT correspondence was first studied
by Polchinski, Susskind and Toumbas \cite{precursor}, who found that 
the causal
propagation of classical wave packets in the AdS bulk leads to the
prediction of extremely unusual degree of freedom in the dual gauge theory
description. Causality in the bosonic lightcone string field theory
formulation was first addressed by Martinec \cite{mar1}, who used the
condition of local commutativity for the string fields to define a
lightcone for the string.

String field theory holds the promise of giving a nonperturbative
formulation of string theory beyond the usual first quantized form.
This would be invaluable to the determination of the vacuum and to the  
formulation and discovery  of the hidden string symmetry which give 
string theories their finiteness and other unusual properties, e.g. 
dualities. 
Lightcone string field theory has the advantage of being
manifestly unitary and also that interactions are 
local in the lightcone time $x^+$. Hence a conventional 
canonical quantization can be performed and a second
quantized operator formalism exists for 
the interacting theory \footnote{
The canonical quantization of the covariant open bosonic 
string field theory  was constructed in \cite{maeno}, 
and more recently  by Erler and Gross\cite{EG}. The later paper also
discussed the issues of locality and causality in the framework of the
covariant string field  theory.
}. 
The basic object in string field theory is the string field operator
$\Phi$ that creates or annihilates string. Observables in the theory
are expressed in terms of $\Phi$. For example the bosonic part of the 
free string lightcone Hamiltonian is given by
\be
H_2 = \frac{1}{2} \int d p^+ \cD^8 P(\s)  \;
\Phi^\dag P^2  \Phi.
\ee 

By employing a 
local commutativity condition on the lightcone string field,
Martinec  constructed the string lightcone for   string theory in
flat spacetime. Imposition of   
such a condition is quite natural since a string field can be written  
as a sum of all the component fields in the theory 
and it is reasonable to impose the condition of 
local commutativity  on the component fields. 
The result of \cite{mar1} gives a natural definition of what may be called 
the string lightcone. It is a natural extension of the usual 
particle lightcone and includes the higher string modes contribution.
The result of \cite{mar1} was later generalized in the 
framework of covariant string field theory \cite{hata-oda}. The
{\it same} result is obtained:
\be \label{slc-flat}
\mbox{the two string fields commute if} \quad 
\int_0^\pi d \s (X^\mu(\s) - Y^\mu(\s))^2 >0.
\ee
This strongly suggests that concept of string local commutativity and 
the physics that one can extract from the
string field commutator is gauge invariant.

These calculations were performed at the free string level. In the
case of quantum field theory of point particles,
interaction is local and the local commutativity condition (hence the
lightcone) is not modified by the interaction. 
The corresponding situation in the case
of string theory  is less clear because, although 
string interact locally via splitting and jointing, 
the string itself is extended. Effects of string interaction on the 
local commutativity condition was first studied by Lowe,  
Susskind and Uglum \cite{lsu}. These authors found that, 
due to  the 3-string interaction, the string field commutator 
ceases to vanish  outside the free string lightcone .  
This result does not necessarily imply that the string theory is
acausal. As these authors argued, this is rather due to the nonlocal nature
of the employed variable: the string field. Later Lowe, Polchinski,
Susskind, Thorlacius and Uglum \cite{lpstu} 
argued that this nonlocal effect could lead to a break
down of the usual nice slice assumption for the low energy 
theory in string theory and proposed that this may lead to a
resolution of the blackhole information paradox. 

A remark concerning the relation between local
commutativity and causality is in order. 
In quantum field theory, one can show that the local (anti)commutativity of
quantum fields (plus other conditions) leads to a causal S-matrix. In
string theory, the connection between string local commutativity and
causality displayed in the S-matrix is however much less clear. In fact
from the above results of \cite{lsu,lpstu}, it was suggested that the reason
for the nonvanishing contribution to the string field commutator was a
effect of nonlocality of strings, rather than acausal behavior of the
theory. The relation between string local commutativity
and string causality was also questioned in the recent paper \cite{EG}.
We find it reasonable and important to make distinction of the
concept of {\it string local commutativity} and the usual concept of
S-matrix causality. In this paper, we will be mainly 
concerned with the local commutativity condition and leave the
important question of how it may be related to  string causality for
the future. 

In the last two years, 
string theory in pp-wave background have been studied with
immense interests, largely due to the remarkable proposal of
Berenstein, Maldacena and Nastase (BMN)\cite{BMN}
which states that a sector of the SYM operators with large $R$-charge
is dual to the IIB string theory on a pp-wave background. The string
background consists of a  plane wave metric  ($d=10$)
\be 
ds^2 = -2 dx^+ dx^- - \mu^2 \sum_{i=1}^{d-2} (x^i)^2 dx^+ dx^+ 
+ \sum_{i=1}^{d-2} dx^i dx^i, 
\label{pp-wave metric} 
\ee 
together with a RR five-form field and a constant dilaton. 
The metric is invariant under
$SO(8)$ rotation of the transversal coordinates $x^i$. The RR background
breaks it down to $SO(4) \times SO(4) \times Z_2$, where $Z_2$
exchanges the two $SO(4)$ factors.  
Remarkably, even in the presence of  curvature and 
a RR 5-form flux, the string theory is exactly solvable \cite{quan} 
in the lightcone gauge. 
It is therefore an interesting question to ask whether and
how the causal structure of the theory is different from that in 
the flat case. 
Note that asymptotic state is not defined in the pp-wave string theory
due to the increase of curvature at large distance, see \eq{pp-wave metric}. 
Thus the notion of a string S-matrix is not well defined 
and one cannot formulate the condition of 
causality in terms of the analyticity of the S-matrix. However the
local commutativity condition is still  well defined 
and appears to be more fundamental and universal.
Therefore it is interesting to see whether 
the nonvanishing  modification of the string field commutator 
persists in the pp-wave case.  
In our previous paper \cite{ck1}, we used the local 
commutativity condition of the free theory to define 
the string lightcone in the pp-wave string theory. We found that 
two strings in the pp-wave background has vanishing commutator if the condition
Eq.\eq{slc} is satisfied.
This result is a modification of the flat space one. In particular it
reduces smoothly to the flat space one in the flat space limit $\mu
\to 0$ of the background. 
In this paper, we
will study the effect of string interaction on the string local
commutativity.

Important remarks on 
the 3-string vertex in the lightcone pp-wave string theory and its
construction are in order. 
The 3-string vertex in the
lightcone pp-wave string theory has been a subject of interest in the
last two years. Motivated by the AdS/CFT duality, it is natural to
suspect the existence of a correspondence 
(see \cite{3point1,3point2,3point3,3point4,3point5} for some different proposals)
between the three point
functions of the BMN operators and the 3-string interaction vertex in
the pp-wave lightcone string theory. 
Besides the applications, the string vertex is a string theoretic object that
is of interest by itself. The bosonic part of the vertex can be determined uniquely 
by  imposing the continuity condition on the embedding of
the string worldsheet into spacetime  \cite{vs,ari}, or by using a
path integral approach \cite{russo-review}.
We will  review the result in section 2 below. We note that this
part of the vertex is continuous in the $\m\to 0$ limit. 
The construction of the fermionic part of the vertex is however more subtle.
In the flat case, the imposition of the kinematical
symmetries fix the vertex up to a prefactor (which is a polynomial  
in $p^+$), which is then fixed by
the imposition of the dynamical supersymmetries. However in the
pp-wave case, the presence of the $Z_2$ symmetry
leads to two possibilities in the choice of the fermionic vacuum
and in the choice of the fermionic zero modes \cite{z2-1,z2-2}. 
Hence one can construct
two inequivalent fermionic vertices that satisfy the fermionic
kinematical constraints. Both can be completed supersymmetrically 
\cite{vs,ari,paolo} and this had leaded to two different possible candidates 
for the lightcone 3-string vertex in the pp-wave background.
 
The main difference  between these two approaches is  in the 
symmetries that are respected by the vertex \cite{russo-review}. 
In the construction of \cite{z2-1,z2-2,paolo}, 
the $Z_2$ symmetry is realized explicitly.  The resulting vertex is,
however, not continuous in the $\m\to 0$ limit. This is not surprising
since the symmetry of the background is
discretely different from the flat space one. 
In the construction of \cite{vs,ari} the  continuity of interaction
vertex in the $\m\to 0$ limit is required instead. We will refer to
this vertex as the ``$\mu$-continuous vertex''.  This condition
seems natural since the string background is smooth in the $\m\to 0$
limit. As it turns out, this requirement forces a
different choice of the fermionic vacuum and  the fermionic zero
modes \cite{z2-1,z2-2} and hence a  different vertex.  There seems to be
no first principle to fix the interaction directly within the
lightcone framework. 
However, it turns out that
the explicit form of the bosonic part of the vertex is sufficient for
the purpose of our calculation. The reason is the following. 
Completing the vertex to the full supersymmetric case requires the
inclusion of the fermionic vertex and the prefactor. However
as we will explain in section 3, both of these
won't modify our result since their contributions are sub-dominant.
Therefore our results obtained using the  bosonic string vertex
\eq{H3-oscil} is unambiguous and applies to both the $Z_2$-invariant
vertex and the $\mu$-continuous vertex.   
Our main motivation was to use this to compute the effect of the
pp-wave string interaction on the condition of local commutativity of
the lightcone string fields and to see whether this effect is continuous
in the  $\m\to 0$ limit. 

Our result is surprising. We find that, unlike the flat case, the
string field commutator does not receive any  
contribution for strings at causally
disconnected region as determined by the free string lightcone
\eq{slc}. Thus the pp-wave string lightcone is unaffected by the string
interaction!
Intuitively this result could be understood since, 
compared to the flat case, the
pp-wave string theory is more confined and more local 
due to the harmonic oscillator potential arises from the
background.  
Since our result is for any $\mu \neq 0$, together with 
the results of \cite{lsu}, it means that
the effect of the string interaction on the string field commutator is
not smooth in the $\m\to 0$ limit. We remark that   it does not matter which
vertices we use, any linear combination of the $Z_2$-invariant vertex
and the $\mu$-continuous vertex will give the same result.
Our result shows that the tree level pp-wave string theory is not
smoothly connected with the flat space theory  even if the vertex is taken to be so.
Since in any case quantities that one can compute from the theory are
not necessary to be continuously connected with the corresponding quantities
in the flat string theory, there is no compelling reason to require that
the vertex to be continuously connected with the flat 
one\footnote{In fact, recently 
Dobashi and Yoneya \cite{3point5}
proposed that the pp-wave string vertex that is relevant for the
holographic pp-wave/SYM correspondence
is given by  the equal weighted sum of the $Z_2$-invariant vertex
and the $\mu$-continuous vertex. 
As such, the vertex is neither continuous at $\mu=0$ nor $Z_2$-symmetric.
We refer the reader to the end of section 3 for more discussions.
}. 

The organization of the paper is as follows. In section 2, we review
the construction of the string field the pp-wave lightcone string
field theory. We  also construct the 3-string lightcone
interaction in  the number basis. 
In section 3, using the technique of contour deformation, 
we calculate the string field commutator in the presence
of the 3-string interaction. 
We  find that   the string interaction does not modify the string field
commutators in the pp-wave case. We discuss the results and its
relevance and implications in the pp-wave/SYM correspondence .  
We end with a couple of further comments and discussions in section
4.

{\bf Note added: }
While the paper is being typed up, the paper \cite{EG} appeared on the archive
which emphasized and examined the issue of locality and causality 
in string theory from the framework of covariant open string field
theory, and  partially overlaps with our work.

\section{The 3-String Interaction in PP-wave Lightcone String Field Theory }

In this section, we briefly review the construction  of 
the lightcone string field  in the pp-wave background \cite{ck1} and 
construct the 3-string interaction vertex. We will leave $\a'$ arbitrary. Our 
final result  is independent of $\a'$.

\subsection{Lightcone String Field}

The type IIB pp-wave background consists of the metric 
\eq{pp-wave metric}, RR-form and a constant dilaton.  
Fixing the lightcone gauge 
$X^+(\s,\t) =\t$, the bosonic  part of the  string action is 
\be 
S = \frac{1}{4 \pi \a'} \int d \t \int_0^{\pi |\a|} d \s 
[(\del_\t X^i)^2 - (\del_\s  X^i)^2 -\mu^2 (X^i)^2], 
\ee 
where $\a = \a' p^+$. $p^+$ is the lightcone
momentum and is positive (negative) for an outgoing(incoming) string. 
In the following analysis, we will focus on the
bosonic part of the theory. Including the fermionic contribution will 
not modify our conclusion. 
For simplicity, we will often  suppress the transverse indices 
$i$.

The mode expansions of the string coordinates and the conjugate momentum  
$P^i= \del_\t X^i /2\pi \a'$ are given by 
\be 
X^i(\s) = x_0^i + 
\sqrt{2} \sum_{l=1}^\infty x_l^i \cos(\frac{l \s}{\a}), \quad
P^i(\s) =  \frac{1}{\pi |\a|} \big[ p_0^i + \sqrt{2} \sum_{l=1}^\infty
  p_l^i \cos(\frac{l\s}{\a})\big] ,
\ee 
where $0\leq \s \leq \pi |\a|$. 
In terms of the modes, the free string lightcone Hamiltonian is 
\be 
H = \frac{\a'}{|\a|}\sum_{l=0}^\infty \Big(- \frac{\del^2}{\del x_l^2} 
+ \frac{1}{4\a'^2} \o_l^2 x_l^2 \Big), \qquad \o_l=
\sqrt{l^2 + (\mu \a)^2}. 
\ee 
This corresponds to a collection of simple harmonic oscillators with 
frequencies $\o_l/ |\a|$  and masses $ |\a|/(2\a')$. 
The string field solves the Schrodinger equation 
$i \del \Phi /\del x^+ = H \Phi $ and is given by 
\be 
\label{SF-X} 
\Phi(\t,x^-,\vX (\s)) = \int_0^\infty 
\frac{dp^+}{\sqrt{2\pi p^+}} \sum_{\{\vn_l\}} 
A\left( p^+, \{\vn_l\} \right) e^{-i \left( x^+ p^- + x^- p^+ \right)} 
f_{\{\vn_l\}}(\vx_l) + h.c.\; ,  
\ee 
where the coordinate space wave function is defined by 
\be 
f_{\{\vn_l\}}(\vx_l)  :=  
\prod_{l=0}^\infty \varphi^l_{\{\vn_l\}} ( \vx_l ) 
\ee 
with
\be 
\varphi^l_{\{\vn_l\}} ( \vx_l ):= \prod_{i=1}^{d-2} H_{\{n^i_l\}} 
\left( \sqrt{\omega_l/2\a'} x^i_l \right) 
e^{-\omega_l ({x^i_l})^2 / {4\a'}} 
\sqrt{ \frac{ \sqrt{\omega_l/2\a'\pi}}{2^{ n^i_l} (n^i_l!)}} 
\ee 
for each fixed $l$.
$\Phi$ has eigenvalue 
\be H = 2p^- = \frac{1}{|\a|}
\sum_{i=1}^{d-2} \sum_{l=0}^\infty n^i_l \omega_l  .
\label{mass-shell} 
\ee 
The Equal Time Commutation Relation for the string field is 
\be 
\left[ \Phi(x^+,x^-,\vX (\s) ),\; 
\Phi (x^+,y^-,\vY (\s) )\right]  = \delta(x^- - y^-) 
\prod_{i=1}^{d-2} \delta\left[ X^i(\sigma) - Y^i(\sigma)\right]. 
\label{ECTR} 
\ee 
This give rises to the commutation relation for the string
creation-annihilation operators 
\be 
\left[ A\left(p^+, \{n^i_l\} \right), 
A^\dag ( q^+, \{m^j_k\} ) \right] = p^+ \delta(p^+ - q^+) 
\delta_{ \{n^i_l\},\{m^j_k\}}. 
\label{cm relations} 
\ee

By considering the unequal time commutator of the string fields 
$\Phi(x^+,x^-,\vX(\s))$ and  $\Phi(y^+,y^-,\vY(\s))$ , one can 
define the string lightcone \cite{mar1}. 
For the pp-wave case, we found that \cite{ck1} the two string fields 
commute if \footnote{
Here we correct a typo in the final result of the paper \cite{ck1}. 
$\mu$ there should be replaced by $\mu/2$.  
}
\be \label{slc}
\Delta x^-   - \frac{\mu}{4 \sin \left( \frac{\mu}{2} \Delta x^+
\right)}  \sum_{l=0}^\infty
\left[ \left( \vx_l^2 + \vy_l^2 \right) \cos \left( \frac{\mu}{2}
\Delta x^+ \right) - 2 \vx_l \cdot \vy_l
\right] < 0.
\ee
Here $\D x^+ := y^+ -x^+$, $\D x^- := y^- -x^-$. 
This result, when restricted to the zero mode sector,
agrees precisely with the particle lightcone, either derived from the local
commutativity condition of the point particle quantum field theory, 
or from the
geodesic distance from \eq{pp-wave metric}. We remark that the
above obtained zero-mode lightcone also agrees with the one obtained
from the wave propagation point of view using the 
scalar \cite{prop1}, spinor and vector propagator \cite{prop2} 
in the pp-wave background. 

 
\subsection{3-String Interaction}

Consider the interaction of 3 open strings with 
lightcone momentum $p^+_{\sr}$, $ \sr=1,2,3$ respectively.  
The string vertex is required to satisfy all the kinematical 
and dynamical symmetries of the theory. This can be easily achieved
by imposing the corresponding continuity condition. It is convenient 
to use a momentum representation of the string field.
It is given by the Fourier 
transform with respect to $y^-$ and $\vY$:
\be 
\label{SF-P} 
\Phit(x^+,\a, \vP(\s)) := \int dy^- \cD\vY(\s) \;  
e^{i y^- p^+ -i \int \vY \cdot \vP } \Phi(x^+,y^-, \vY(\s)) . 
\ee 
In this basis, the bosonic part of the
three-string interaction Hamiltonian is given by 
\be 
\label{3-string Hamiltonian} 
H_3 = g \int \prod_{{\sr}=1}^3  d \a_\sr \mathcal{D}\vP_\sr(\s) \;
\tilde{h}(\a_\sr,\vP_\sr(\s)) \; 
\prod_{{\sr}=1}^3 \Phit(x^+,\a_\sr, \vP_\sr(\s)) . 
\ee 
Here $g$ is the string coupling, 
$\vP_\sr(\s)$ is the transverse momentum field of 
the $\sr$-th string, and 
The integration measure is defined in terms of the modes 
$\cD\vY(\s) := \prod_{l=0}^\infty d\vy_l$
and $\int \vY \cdot \vP = \sum_{l=0}^\infty \vy_l \cdot \vp_l$. 
Without loss of generality, we assume $\a_1, \a_3>0,
\a_2<0$, and the measure factor $\tilde{h}(\a_\sr, \vP_\sr(\s))$ 
is given by\footnote{
The string with $\a$ which is opposite in sign to the other two's has the 
widest transverse extension and hence the form of the delta-functional.
}
\be 
\tilde{h}(\a_\sr, \vP_\sr(\s)) = \d(\sum \a_\sr)
\int \prod_{\sr=1}^3\cD\vY_\sr(\s) \; 
e^{ i\int \vP_\sr \cdot \vY_\sr} \d(\vY_2 - \vY_1 - \vY_3), 
\ee 
which is basically a continuity condition. 

To pass to the oscillator number basis, we substitute 
\eq{SF-P} and \eq{SF-X} and obtain 
\be 
\label{H3-oscil} 
H_3 = g \int \prod_{{\sr}=1}^3  \frac{d \a_\sr}{\sqrt{2 \pi |\a_\sr|}}
\sum_{\{\vn_{\sr, l}\}} \Vt(\a_\sr, \{\vn_{\sr, l}\}) \prod_{{\sr}=1}^3  
A\left( p_\sr^+, \{\vn_{\sr, l}\} \right) + \cdots\; ,
\ee 
where the $\ldots$ are terms of the form $A A A^\dag, A A^\dag A^\dag, 
A^\dag A^\dag A^\dag $. For the calculation to be performed below, 
only the $A A A$ term is relevant. In \eq{H3-oscil}, 
$\Vt(\a_\sr, \{\vn_{\sr l}\})$ is the 3-string vertex in the 
oscillator number basis. It is given by 
\be \label{3-string vertex} 
\Vt(\a_\sr, \{\vn_{\sr, l}\}) = \int
\prod_{{\sr}=1}^3 \prod_{l=0}^\infty d\vp_{\sr, l} \; 
\tilde{h}(\a_\sr,\vP_\sr(\s)) \prod_{{\sr}=1}^3  \ft_{\{\vn_{\sr, l}\}}
(\vp_{\sr, l}) ,
\ee 
where 
\be \ft_{\{\vn_l\}}(\vp_{l}) := 
\int\prod_{l=0}^\infty d \vy_{l} \; e^{-i \sum_l \vy_l \cdot \vp_l}
f_{\{\vn_l\}}(\vy_{l}) 
\ee 
is the momentum space wave function. 
We note that this part of the vertex is continuous 
in the $\m\to 0$ limit. 

The bosonic vertex \eq{H3-oscil} should be completed 
with the fermionic vertex and the 
prefactor in order to respect fully both the kinematical and the dynamical
symmetries of the theory. 
As mentioned in the introduction,  
the construction of the fermionic part of the vertex is more subtle.
It has been pointed out that due to the presence of the 
$Z_2$ symmetry, two possible choices in the fermionic vacuum
and in the choice of the fermionic zero modes \cite{z2-1,z2-2} 
is allowed. Hence one can construct
two inequivalent fermionic vertices that satisfy the fermionic
kinematical constraints. Moreover both can be completed supersymmetrically 
\cite{vs,ari,paolo} and this leads to two different possible candidates 
for the lightcone 3-string vertex in the pp-wave background. 
These two vertices have the same bosonic part, but different fermionic parts
and different prefactors. However it is easy to see that
for our computation, it is enough to use the above constructed 
bosonic vertex. 
One can easily see that \cite{Lowe} 
the fermionic contribution is sub-dominant in the $p^+ \to \infty$ limit,
as compared to the
exponentially  growing behavior of the bosonic contribution (see
\eq{K-int-final} below). This is also the same for the
contribution of the prefactor since  the prefactor (in both cases) is 
a polynomial in $p^+$. Hence the bosonic contribution dominates in our
computation. 


\section{String Interaction and String Field Commutator}

In this section, we investigate the effects of string interaction on the 
local commutativity condition of the string fields. Our analysis follows 
closely that of \cite{lpstu}. 

Consider the amplitude 
\be \label{m-elt}
M = {}_{\rm H}\lvac \left[ \Phi_H(1), \Phi_H(2) \right] |3\rangle_{\rm H},  
\ee 
of two string fields,
denoted by 1 and 2, with a \third spectator state. 
The spectator state  is necessary for a possible non-zero contribution at
first order in the string coupling constant $g$. 
The subscript ${\rm H}$ means that everything is in the Hamiltonian picture.
Passing to the interaction picture, we have   
\be 
M = \langle 0;x^+_1| \Phi_I(1) U_I(x^+_1, x^+_2) \Phi_I(2) |3;
x^+_2 \rangle - \{1 \leftrightarrow 2\} ,
\ee 
where
$U_I(x^+_1, x^+_2)$ 
is the time evolution operator in the
interaction picture. In the leading order of string coupling, 
it is
\be 
U_I(x^+_1, x^+_2) 
= 1 + ig \int_{x^+_1}^{x^+_2} dx^+ H_3(x^+) + \cdots \;. 
\ee 
Hence up to first order in $g$ we have 
\be 
M = M^{(0)} + M^{(1)}, 
\ee 
where the zeroth order amplitude 
\be 
M^{(0)} = \lvac \left[ \Phi_I(1), \Phi_I(2) \right]|3 \rangle
\ee 
is a matrix element of the commutator of the two string fields, 
and 
\be 
M^{(1)} = ig\int_{x^+_1}^{x^+_2} dx^+ \lvac \Phi_I(1) H_3(x^+) \Phi_I(2) 
+ \Phi_I(2) H_3(x^+) \Phi_I(1) - \Phi_I(1) \Phi_I(2) H_3(x^+) |3\rangle .
\ee 
For strings outside the string lightcone \eq{slc},
we see immediately that $M^{(0)} = 0$. 
Any possible causality violations will come from a non-zero $M^{(1)}$.

Now, since $H_3$ is of the form $\Phi^3$
and the string field of the form $\Phi \sim A + A^\dag$, 
we can break the interaction vertex down to terms with equal 
number of creation and annihilation operators,  $H_3 = H_{3\text{aaa}} + 
H_{3\text{aac}} + H_{3\text{acc}} + H_{3\text{ccc}}$. 
It is easy to see that unless the spectator state is a single string
state of the form 
$|3\rangle = A^\dag(p^+_3, \{\vn_{3,l}\})\vac$, 
$M^{(1)}$ will be identically zero. With this choice
for the spectator state, we have  
\be 
M^{(1)} = ig\int_{x^+_1}^{x^+_2} dx^+ \lvac \Phi_\text{a}(1)
H_{3\text{aac}}(x^+) \Phi_\text{c}(2) + 
\Phi_\text{a}(2) H_{3\text{aac}}(x^+) \Phi_\text{c}(1) - \Phi_\text{a}(1)
\Phi_\text{a}(2) H_{3\text{acc}}(x^+) |3\rangle .
\ee 
Substituting \eq{SF-X} and \eq{H3-oscil}, it is easy to obtain
\bea \label{M1-ff}
M^{(1)} && =  ig \int_{x^+_1}^{x^+_2} d \t
\prod_{\sr=1}^2 \int_{-\infty}^\infty  
\frac{d\a_\sr}{\sqrt{2\pi|\a_\sr|}} F(\a_1, \a_2)  \nn\\
&& \cdot \sum_{\{\vn_{1,l}\}, \{\vn_{2,l}\}} 
\left( f_{\{\vn_{1,l}\}}(\vx_{1,l}) f_{\{\vn_{2,l}\}}
(\vx_{2,l}) \Vt(1,2,3)\; 
e^{- i\t \sum_{\sr=1}^3 p^-_\sr} 
e^{i \sum_{\sr=1}^2 p^-_\sr x^+_\sr +  p^+_\sr x^-_\sr }
\right), \;\;\;
\eea
where the function $F$  is defined by
\be
 F(\a_1, \a_2) := \Theta(\a_1) \Theta(-\a_2) 
+ \Theta(-\a_1) \Theta(\a_2) - \Theta(-\a_1) \Theta(-\a_2) .
\ee 
Using \eq{3-string vertex} for the explicit expression of $\Vt(1,2,3)$,
and using sum rule for the Hermite polynomial, 
one can easily calculate the sum in the second line of \eq{M1-ff} 
and obtain
\bea \label{M1-JJ}
M^{(1)} && =  ig \int_{x^+_1}^{x^+_2} d \t
\int_{-\infty}^\infty
\frac{ d\a_1}{2\pi\sqrt{|\a_1(\a_1 + \a_3)|}} \nn\\
 &&  \cdot \int \prod_{\sr=1}^3 \prod_{l=0}^\infty d\vy_{\sr,l} \; 
\d (\vY_2 - \vY_1 - \vY_3 ) 
J_{1,l}(\a_1, \vx_{1,l}, \vy_{1,l}) J_{2,l}(\a_2,\vx_{2,l},\vy_{2,l}) \nn\\
&& \cdot f_{\{\vn_{3,l}\}}(\vy_{3,l}) 
e^{- i \t p^-_3} e^{-i p^+_3 x^-_2} e^{-ip_1^+ \D x^-} .
\eea
Here we have introduced the shorthand notation 
\bea 
J_{\sr,l}(\a_\sr, \vx_{\sr,l}, \vy_{\sr,l}) & := & 
\left(\frac{\o_{\sr,l}/2\a'}{\pi} 
\frac{1}{1 - e^{-i \t_\sr \o_{\sr,l}/|\a_\sr|}}\right)^{(d-2)/2} \nn \\ 
& &  \cdot \exp \left\{ \frac{\o_{\sr,l}/2\a'}{2i\sin\left(\frac{\t_\sr
\cdot \o_{\sr,l}}{2|\a_\sr|} \right)} \left[2 \vx_{\sr,l} \cdot \vy_{\sr,l} - 
\left( {\vx_{\sr,l}}^2 + {\vy_{\sr,l}}^2 \right) 
\cos \left( \frac{\t_\sr \o_{\sr,l}}{2|\a_\sr|} \right) 
\right] \right\} \;\;\;.
\eea 
and $\D x^- := x^-_2 - x^-_1$ .
The $\t$-dependence enters through $\t_\sr := \t - x^+_\sr$. 
For the kinematic situation 
we are considering here ($\a_1, \a_3 >0, a_2<0$), the delta-functional is 
\be \label{delta}
\d(\vY_2 - \vY_3 - \vY_1) = 
\prod_{m=0}^\infty \d\left( \vy_{2,m} 
- \sum_{n=0}^\infty \big (\left| \frac{\a_3}{\a_2} \right| X_{mn}^{(3)} \vy_{3,n} 
+ \left| \frac{\a_1}{\a_2} \right| X_{mn}^{(1)} \vy_{1,n}\big) \right). 
\ee
The definition and properties of the matrices $X$ are recalled  
in the appendix. Also for our case  $F(\a_1,\a_2) =1$.

To proceed further, one may  write  $M^{(1)}$ in the form
\be \label{MK}
M^{(1)} = \int_{-\infty}^\infty d \a_1 K(\a_1) \; e^{- i  \a_1 \D x^-  /\a'  },
\ee
where 
\bea \label{K-gen}
K(\a_1) && := \frac{ig}{2\pi\sqrt{|\a_1(\a_1 + \a_3)|}} 
\int_{x^+_1}^{x^+_2} d\t
\int \prod_{\sr=1}^3 \prod_{l=0}^\infty d\vy_{\sr,l} 
\d (\vY_2 - \vY_1 - \vY_3 ) \nn\\
&& \cdot \prod_{l=0}^\infty 
J_{1,l}(\a_1, \vx_{1,l}, \vy_{1,l}) J_{2,l}(\a_2,\vx_{2,l},\vy_{2,l}) 
\cdot f_{\{\vn_{3,l}\}}(\vy_{3,l})  e^{- i \t p^-_3} e^{-i p^+_3 x^-_2} .
\eea
Now let us focus our attention on the $\a_1$ integral. As was done 
in \cite{mar1,ck1}, we can write $M^{(1)}$ as 
\be 
\int_0^\infty d\a_1 K(\a_1) e^{- i  \a_1 \D x^-  /\a' } 
+ \int_0^\infty d\a_1 K(-\a_1) e^{ i  \a_1 \D x^-  /\a'} .
\ee 
Rotate the first integral by sending $\a_1 \rightarrow i\a_1$ 
and the second term by sending $\a_1 \rightarrow -i\a_1$. 
Then 
\be \label{M1-rot}
M^{(1)} = i \int_0^\infty d\a_1 K(i\a_1) e^{\a_1 \D x^-  /\a'} 
- i \int_0^\infty d\a_1 K(i\a_1) e^{\a_1 \D x^-  /\a' } .
\ee 
If each individual integral converges, the two terms cancel each other 
and hence $M^{(1)} =0$.
For that, we must examine the large $\a_1$ behavior of $K(i\a_1)$.
In \cite{lsu,lpstu}, 
it was found that for the flat case, the integral does not converge 
and the integral could 
picks up contribution from region outside the free string lightcone. 
Their result demonstrates the break down of local commutativity in the 
lightcone theory. However, it does not necessary mean that  
causality is violated. As the authors argued, this is rather due to the
nonlocal nature of string itself.
Below we will examine the same issue in the case of pp-wave string theory. 

The above analysis was carried out for 
the general case with arbitrary string fields. 
It will be illuminating to consider a simplified situation
where the $1^\text{st}$ and $2^\text{nd}$ string fields are taken to be
the lowest component fields with:
\be 
\label{restricted ns} 
\vn_{1,l} = \vn_{2,l} =
\begin{cases} 0, & \text{when $l \geq 1$}, \\ 
\text{arbitrary}, & \text{when $l=0$}. 
\end{cases} 
\ee 
The component field is obtained by integrating the 
string field with
$\prod_{l=1}^\infty d\vx_{l} \varphi^l_{\{\vn_{l}\}}(\vx_{l})$. 
This gives 
\be
T(\t,x^-,\vx) = \int
\frac{dp^+}{\sqrt{2\pi p^+}} \sum_{\vn_0}
a ( p^+,\vn_0) e^{-i \left( x^+ p^- + x^- p^+ \right)}
\varphi^0_{\{\vn_0\}}(\vx) + h.c. \, ,
\ee
where we have defined 
$ a ( p^+,\vn_0):= A\left( p^+, \vn_0, \{\vn_{l\geq 1} =0 \} \right)$
and in the following we often denote the zero mode $\vx_0$ by $\vx$ 
for simplicity.
Furthermore, we restrict the \third string to be the following 
spectator state:
\be 
|3\rangle = A(p^+_3, \{\vn_{3,l}\})\vac , \quad \mbox{with}\quad
\vn_{3,l} = 0, \; \mbox{for all} \ l .
\ee 
We note that  $p^-_3 = 0$. 

Following the same procedures as above, 
it is easy to obtain \eq{MK} with $K(\a_1)$ now taking the form
\bea
&& K(\a_1) =  \frac{ig \; e^{-i p_3^+ x_2^- }}{2\pi\sqrt{|\a_1(\a_1 + \a_3)|}} 
 \int_{x^+_1}^{x^+_2} d \t
\int \prod_{\sr=1}^3 \prod_{l=0}^\infty d\vy_{\sr,l} 
\d (\vY_2 - \vY_1 - \vY_3 ) \nn\\
&& 
\;\;\cdot J_{1,0}(\a_1, \vx_{1,0}, \vy_{1,0}) J_{2,0}(\a_2,\vx_{2,0},\vy_{2,0}) 
\prod_{l=1}^\infty \varphi^l_{\{0\}}(\vy_{1,l})\varphi^l_{\{0\}}(\vy_{2,l})
\cdot f_{\{0\}}(\vy_{3,l})
 . \;\;\;\;\;\;
\eea 
We note that, compared with \eq{K-gen}, 
the product $\prod_{l=1}^\infty J_{1,l}(\cdot)  J_{2,l}(\cdot)$ 
in the second line there is replaced by 
$\prod_{l=1}^\infty \varphi^l_{\{0\}}(\vy_{1,l})
\varphi^l_{\{0\}}(\vy_{2,l})$ above due to the condition \eq{restricted ns}.
Now we perform the contour rotation and focus on the integrals of the $\vy$'s. 
Let us first intergate $d\vy_{2,l}, l\geq 1$ using the nonzero mode 
delta functions.
One can show that the resulting integral of $\vy_{1,l}$ and 
$\vy_{3,l}$, $l\geq 1$ 
is independent of the zero modes $\vy_{1,0}$ and $\vy_{3,0}$ in the large
$\a_1$ limit and  so can be calculated easily. Next we
integrate out $d\vy_{3,0}$ using the zero mode 
delta function. Therefore 
in the  large $\a_1$ limit,  
\be \label{K-large-a}
e^{ip^+_3 x^-_2} \frac{K(i\a_1)}{ig} \sim \int_{x^+_1}^{x^+_2} d\t
\int d\vy_{1} d\vy_{2}\; \Jh_{1,0} \Jh_{2,0} 
\exp \left[ + \frac{\o_{3,0}}{4\a'} \left( \left|
\frac{\a_2}{\a_3} \right| \vy_2 - \left| \frac{\a_1}{\a_3} \right| 
\vy_1 \right)^2 \right] 
\ee
up to an unimportant $\a_1$-dependent proportional factor which is 
sub-dominant in large  $\a_1$ limit. Here we have 
denoted $\vy_{\sr,0}$ by $\vy_{\sr}$ for simplicity.
Also we have used the hat $\hat{}$ 
to denote the corresponding quantities with the substitution 
$\a_1 \to i \a_1$. For example, $\omh_{1,l} = \sqrt{l^2- \m^2 \a_1^2}$
in $\Jh_{1,0}$.
After the contour rotation and taking the large $\a_1$ limit, we have  
\be 
\hat{\a_2} \sim -i \a_1, \quad \mbox{and}\quad 
\omh_{1,l}, \; \omh_{2,l} \sim i \m \a_1 .
\ee 
Now $\Jh_{\sr,0}$ takes the form
\be
\Jh_{\sr,0} \sim  \exp\left( 
-A_\sr \left(\vx_\sr^2 + \vy_\sr^2 \right) + 2 B_\sr \vx_\sr \cdot \vy_\sr
\right)
\ee 
with
\be 
A_\sr = \frac{\m \a_1/(2\a')}{2\tan (\frac{\m}{2}|\t_\sr|) } , \quad
B_\sr = \frac{\m \a_1/(2\a')}{2\sin (\frac{\m}{2}|\t_\sr|) }
\ee
and thus the $\vy_1, \vy_2$ integral takes the form
$ \int d\vy_1 d\vy_2 \exp(-\sum_{\sr,\ss} N_{\sr \ss} \vy_\sr \cdot \vy_\ss 
+ \sum_{\sr} \vS_\sr \cdot \vy_\sr) $
and  can be easily carried out. We obtain for the 
$\vy_1, \vy_2$ integral in \eq{K-large-a},
\be \label{res-int}
\exp{\left[
-\frac{\m \a_1/(2\a')}{2\tan (\frac{\m}{2}\D x^+) }(\vx_1^2 +\vx_2^2)
+ \frac{\m \a_1/(2\a')}{\sin (\frac{\m}{2}\D x^+) } \vx_1 \cdot \vx_2
\right]}
\ee
in the leading large $\a_1$ limit. It is remarkable that the various 
coefficients of $N_{\sr\ss}, \vS_\sr$ combine to make the result \eq{res-int} 
$\t$ independent. Hence the $\t$ integral in \eq{K-large-a} 
can be calculated trivially. Finally we obtain for \eq{M1-rot}
\bea \label{K-int-final}
&& \int_0^\infty d\a_1 K(i \a_1) e^{\frac{\a_1}{\a'}\D x^-} \nn\\
&&\sim\int_0^\infty d\a_1 \exp{\left[
\frac{\a_1}{\a'} 
\Big( \Delta x^- 
- \frac{\mu}{4 \sin \left( \frac{\mu}{2}\Delta x^+ \right)} 
\big( 
( \vx_1^2 + \vx_2^2 ) \cos \left( \frac{\mu}{2} \Delta x^+ \right) 
- 2 \vx_{1}\cdot \vx_{ 2}
\big)
\Big) \right]} . \;\;\;\;\;\;\;\;
\eea
The exponent of the integrand is precisely the tree level string 
lightcone \eq{slc} restricted to the zero modes. Thus we have shown
that, unlike the flat case, the matrix element \eq{m-elt} does not receive
contribution from region outside the free string lightcone. 

As we remarked above, to be fully supersymmetric, 
the bosonic vertex has to be completed with the fermionic
vertex and a prefactor which is needed for the preservation of the supersymmetries. 
Also the bosonic string field has to be replaced by the 
lightcone string superfield \cite{sft4} so that we commute the matrix
element of the commutator of two string superfields. Now 
the Grassmannian factor  makes sub-dominant
contributions to the contour-deformed integral in the limit $p^+ \to
\infty$ and does not affect  the convergence of the contour-deformed integral.  
This is also the case for the prefactor as it is 
a polynomial in $p^+$. Therefore we 
conclude that the commutator of the string fields 
is unaffected by the pp-wave string interaction.
This is the main result of our paper.

Our result is surprising. Recall that in the flat case,  the
string field commutator was found to receive additional nonvanishing 
contribution \cite{lsu,lpstu} from the interaction 
even if the two strings were outside the
free theory string lightcone of each other. 
Since  the pp-wave background and the bosonic part of the 
vertex  are both continuous in
the $\m\to 0$ limit, one may natively thought that the pp-wave string field
commutator should also receive additional contributions, 
at least in a neighborhood of $\mu =0$. Our result shows that this is
not the case and the matrix element \eq{m-elt} is discontinuous at
$\mu =0$. 
Technically the reason for the discontinuity is because  the
$\mu \to 0$ limit does not commute with the procedure of summing up the
contributions from the infinite tower of string states. 
Our result shows that the UV behavior of the theory is sensitive 
to the IR parameter $\mu$.
This mixing of the IR and UV effects is similar to the IR/UV mixing in the 
noncommutative field theory \cite{IRUV}.

Since the pp-wave lightcone string field theory is  
not continuously connected with the flat space string theory, 
there is no compelling reason to
require  that the 3-string vertex to be continuous at
$\mu =0$. What about the $Z_2$ symmetry? We would like to make the
following remark. 
Without additional input, one cannot fix the form of the
lightcone vertex uniquely from the supersymmetries alone. One may fix
the form of the vertex by requiring the $Z_2$ symmetry. However there is a
possibility that the symmetry is spontaneously broken. Since the
pp-wave background is obtained from the AdS background by performing a
Penrose limit,  a reasonable possibility that may help 
to understand better the pp-wave string
interaction is to try perform this limit carefully on the 
dynamics on the AdS side. 
This interesting idea has been pursued recently by 
Dobashi and Yoneya \cite{3point5}. They
propose that the pp-wave string vertex that is relevant for the
holographic pp-wave/SYM correspondence 
is given by  the equal weighted sum of the $Z_2$-invariant vertex 
and the $\mu$-continuous vertex. It turns out that this particular
combination  coincides with the vertex proposed previously in
\cite{3point4}. These authors also provide some
intuitive understanding of the role of each parts of the vertex: 
the  $Z_2$-invariant vertex describes the ``bare'' interaction, 
while the $\mu$-continuous vertex describes the mixing of the BMN operators. 
Thus according to this proposal, not only the continuity of $\mu$ is
not maintained, also the $Z_2$ symmetry is broken due to the mixing effects.
The breaking of the $Z_2$ symmetry has also been
revealed in previous field theory calculations \cite{Z2-break}.  
In principle, one can fix the form of the lightcone vertex by starting from
the covariant Witten string field theory by doing the lightcone 
gauge fixing. To confirm this breaking from a more fundamental point of
view will be very exciting.

\section{Discussions and Conclusions}
 
Using the framework of lightcone string field theory, 
we computed the effect of string interaction on the
string field commutator. We found that the 
string field commutator is unaffected by the pp-wave string
interaction.  This is the main result of our paper.
Although our computation was performed at the first order in the string
coupling, it is natural to conjecture that this remain the case to all
order. Also for technical reason, we have restricted ourself to compute
the matrix element for the special case \eq{restricted ns} 
where the $\tau$-integral
is managable. The general case \eq{MK}, \eq{K-gen} 
is more complicated. However we expect our conclusion remains the same. 
 

In this paper, we used a lightcone gauge fixed string field theory
framework. Technically, pp-wave string theory is tractable so far 
only in the lightcone gauge\footnote{Using the pure spinors 
formalism, Berkovits \cite{berk} 
has constructed an alternative covariant quantization for the pp-wave string theory.
background. However the presence of a non--trivial
background gives rise to a complicate world-sheet action, and explicit
computations of amplitudes in this framework have not been done yet.}.
It is possible that the conclusion that one can drawn from 
the string fields commutator computation is gauge dependent. 
We don't have a proof that this is not the case. However it has been shown that 
in the case of free string in flat spacetime, the covariant 
string field theory and the lightcone string field theory gave 
the same result: demanding the vanishing of the string field
commutator give rises to the same string lightcone. 
This strongly suggests that the physics that one can extract from the  
string field commutator is indeed gauge invariant. We believe this is the same even 
when interaction is included, i.e. local
commutativity of string fields is a gauge invariant physical concept.
It will be very interesting to determine
the string field commutator in the covariant string field theory
\cite{witten} and see if one obtains the same nonlocal effect as in 
\cite{lsu,lpstu}. The recently developed Hamiltonian formalism \cite{EG} could be
helpful in this respect. 
 
If the string local commutativity is indeed gauge invariant, 
then what is its physical meaning? is it in some way  
related to the causality in string theory? 
It is important to understand what is the physical meaning of 
string local commutativity. Note that the usual operational definition 
of causality  is based on the propagation of light signal and is a
concept relevant for the low energy point particle theory.
Imagine in a short distance scale where string scale is relevant. At
this scale, we cannot ignore the nonlocal nature of the string probe. 
Obviously the usual  definition of causality appears  
insufficient and  we  need a new operational definition. 
An important requirement is that this string form of
causality should reduce to the usual form of causality in the 
low energy limit. We think the 
string local commutativity may have a chance to be related to this string
form of causality\footnote{Indeed if we use the string local commutativity
to define a string lightcone, then this notion of causality indeed
reduce to the ordinary one in the low energy limit since the effect 
found in \cite{lsu,lpstu} vanishes as one takes $\a'
\to 0$.}  . 
It is  an important question to explore. 



In this paper, we examined the causality property in the interacting open
string theory. The closed string case needs more discussion. An
immediate question is what is the meaning of time in quantum gravity.
It is generally believed that spacetime is a
classical concept in a theory of quantum gravity, and the concept of 
spacetime loss its meaning when gravitational field is quantized. One
may therefore question about the meaning of causality in the closed
string theory. Although spacetime itself is not an observable in
quantum gravity, it should still be useful to  impose sensible condition, 
e.g. causality or local
commutativity, in the formulation of the theory.
For example we note the interesting suggestion of Teitelboim
\cite{teitelboim} who proposed to impose the condition of causality on 
the classical configurations (which makes senses) to be integrated 
in the path integral of gravity. 
In this sense, the study of causality become a valid question in
closed string theory. And we propose that  local commutativity
could be a useful characterization for such in the closed string  theory too.

\acknowledgments

CSC would like to thank 
Valya V. Khoze and Rodolfo Russo for helpful comments on the manuscript.
We acknowledge grants and fellowships 
from the Nuffield foundation and EPSRC of UK, the
National Center of Theoretical Science and National Science Council  of Taiwan.

\appendix

\section{Fourier Transform for Open Strings}

Consider three open strings with 
lightcone momenta parametrized by $\a_{\sr}$ and with $\a_1, \a_3>0,
\a_2<0$. 
The momentum of the strings are given by
\be \label{fourier}
P_\sr(\s) =  \frac{1}{\pi |\a|} \big[ p_{\sr,0} + 
\sqrt{2} \sum_{l=1}^\infty  p_{\sr,l} \cos(\frac{l\s}{\a})\big] ,
\ee
with the coordinates of the three strings  parametrized by 
\be
\begin{array}{lll}
 \s_1 = \s, & &0 \leq \s \leq \pi \a_1, \nn\\
 \s_3= \s -\pi \a_1,& & \pi \a_1 \leq \s \leq \pi (\a_1 +\a_3), \nn\\
 \s_2 = - \s, & &0 \leq \s \leq \pi (\a_1 + \a_3).
\end{array}
\ee
The sum $P(\s):= \sum_{\sr=1}^3 P_\sr(\s)$ admits the Fourier decomposition like
\eq{fourier} with 
\be
p_m = \sum_{\sr=1}^3 \sum_{n=0}^\infty X^{\sr}_{mn} p_{\sr,n}, \quad m\geq 0.
\ee
The matrix $X^{(2)}_{mn}=\d_{mn}$ and for $\sr =1,3$
\be
X^{(\sr)}_{mn} = 
\begin{cases}
\Xt^{(\sr)}_{mn}\,,\qquad m>0\,,n>0 \\ 
\frac{1}{\sqrt{2}}\Xt^{(\sr)}_{m0}\,,\qquad m>0 \\ 1\,,\qquad m=0=n,
\end{cases}
\ee
where for $m>0, n\geq 0$,
\be
\Xt^{(1)}_{mn}:= (-1)^n\frac{2m\b}{\pi}\frac{\sin m\pi\b}{m^2\b^2-n^2}\,,\quad
\Xt^{(3)}_{mn}:= \frac{2m(\b+1)}{\pi}\frac{\sin m\pi\b}{m^2(\b+1)^2-n^2}, 
\ee
and $\b = \a_1/\a_2$, $\b+1 = -\a_3/\a_2$.

The matrices $X^{(\sr)}$ satisfy the following identities
\be
(X^{(\sr) T} X^{(\ss)})_{mn} = -\frac{\a_2}{\a_\sr} \d_{\sr\ss}, \quad 
\sr =1,3
\ee
and
\be
\sum_{\sr=1}^3 \a_{\sr} (X^{(\sr)} X^{(\ss) T}) =0.
\ee

\end{document}